\documentclass[aps,pra,superscriptaddress,twocolumn]{revtex4}
\usepackage{color,graphicx,hyperref,ulem}
\usepackage{amsmath, amssymb}
\DeclareMathOperator{\diag}{diag}
\DeclareMathOperator{\sinc}{sinc}
\usepackage{siunitx}

\definecolor{gruen}{rgb}{0,0.7,0.1}
\definecolor{blau}{rgb}{0,0.1,0.7}
\definecolor{rot}{rgb}{0.8,0,0}

\begin{document}

\title{Bloch-Messiah decomposition and Magnus expansion for parametric \\ down-conversion with monochromatic pump}

\author{Tobias Lipfert}\email[]{tobias.lipfert@univ-lille1.fr}
\affiliation{Univ.~Lille, CNRS, UMR 8523 - PhLAM - Physique des Lasers Atomes et Mol\'ecules, F-59000 Lille, France}

\author{Dmitri~B.~Horoshko}\email[]{dmitri.horoshko@univ-lille1.fr}
\affiliation{Univ.~Lille, CNRS, UMR 8523 - PhLAM - Physique des Lasers Atomes et Mol\'ecules, F-59000 Lille, France}
\affiliation{B.~I.~Stepanov Institute of Physics, NASB, Nezavisimosti Ave.~68, Minsk 220072 Belarus}%

\author{Giuseppe Patera}
\affiliation{Univ.~Lille, CNRS, UMR 8523 - PhLAM - Physique des Lasers Atomes et Mol\'ecules, F-59000 Lille, France}

\author{Mikhail~I.~Kolobov}
\affiliation{Univ.~Lille, CNRS, UMR 8523 - PhLAM - Physique des Lasers Atomes et Mol\'ecules, F-59000 Lille, France}

\date{\today}

\begin{abstract}
	We discuss the Bloch-Messiah decomposition for the broadband squeezed light generated by type-I parametric down-conversion with monochromatic pump. Using an exact solution for this process, we evaluate the squeezing parameters and the corresponding squeezing eigenmodes. Next, we consider the Magnus expansion of the quantum-mechanical evolution operator for this process and obtain its first three approximation orders. Using these approximated solutions, we evaluate the corresponding approximations for the Bloch-Messiah decomposition. Our results allow us to conclude that the first-order approximation of the Magnus expansion is sufficient for description of the broadband squeezed light for squeezing values below 12.5 dB. For higher degrees of squeezing we show fast convergence of the Magnus series providing a good approximation for the exact solution already in the third order. We propose a quantitative criterion for this ultra-high-gain regime of parametric down-conversion when the higher-orders terms of the Magnus expansion, known in the literature as the operator-ordering effects, become necessary.
\end{abstract}

\maketitle

\section{Introduction}

	Squeezed states of light are nonclassical states with unique features interesting from both the fundamental and the practical points of view \cite{Dodonov2002,AndersenGML2016,Kolobov1999}. They are typically generated by parametric down-conversion (PDC) and four-wave mixing, and find numerous applications in laser interferometers, including gravitational wave detectors \cite{ChuaSSM2014,Schnabel2017}, in quantum metrology~\cite{WolfgrammCBPKM2010}, and in various protocols of quantum information, from quantum teleportation to quantum computation \cite{BraunsteinLoock2005,Weedbrook2012}. In the latter area of research multimode squeezed states are recognized as a key resource for the  measurement-based continuous variable quantum computation~\cite{LloydB1999,Menicucci2008,Yokoyama2013,MarshallJSGWA2016,DouceMKDCMLF2017}. The efficiency of employing squeezed states depends crucially on the degree of squeezing. As consequence, there is a high demand for squeezed states with highest possible degree of squeezing.

	The continuous-wave (CW) narrow-band squeezed light is, perhaps, the best-known realization of squeezing in optics, and the experiments in this direction reach the record values of 15 dB squeezing in a band of about 100 MHz \cite{Vahlbruch2016}. On the other hand, broadband PDC opens the possibility for generating multiple modes of squeezed light at once.  Broadband squeezing (from GHz to tens of THz) can be obtained in pulsed regime both in single-pass~\cite{YurkeGSP1987,BenninkB2002,WasilewskiLBR2006,Agafonov2010,Allevi2014,Perez2015} or cavity-enhanced configurations~\cite{ValcarcelPTF2006,Patera2010,ShifengTF2012,Pinel2012}. It can also be obtained in CW or quasi-CW regimes of PDC with the use of aperiodically poled quasi-phase-matched crystals~\cite{Horoshko2013,Horoshko2017,Chekhova2018}. In any case, the detection and implementation of broadband squeezed states requires precise definition of the squeezing eigenmodes in order to use the squeezing most efficiently. The mathematical background for this type of modal decomposition is based on the Bloch-Messiah decomposition of the field variables~\cite{BlochM1962,BalainDI1965,ArvindDM1995,Braunstein2005,AdessoRL2014,Ferrini2014,Cariolaro2016a,Cariolaro2016b}. In the low-gain regime of PDC the equivalent procedure is known as the Schmidt decomposition of the two-photon state \cite{LawWE2000,CaspaniBG2010,Horoshko2012,Perina2015}. The transition from the Schmidt modes of the low-gain regime to the squeezing modes in the high-gain regime has been discussed in the literature but is far from being completely understood. A numerical study demonstrated recently \cite{ChristBMS2013} that for relatively moderate squeezing the squeezing eigenmodes are given by the Schmidt modes, but above certain degree of squeezing these two sets are different.
	
	Analytical analysis of squeezing eigenmodes at high gain is complicated by the non-stationarity of the problem. A powerful mathematical tool in this case is known in the literature as the Magnus expansion \cite{Magnus1954,BlanesaCOR2009} of the quantum-mechanical evolution operator. Truncation of this expansion, which we shall call the Magnus approximation (MA), preserves the unitarity of the evolution operator which is crucial for a proper description of the squeezed states of light. The first-order MA is equivalent to disregarding quantum-mechanical effect of operator ordering in the Dyson series, and has been used by several authors for the definition of the squeezing modes at high gain \cite{BenninkB2002,WasilewskiLBR2006,ShifengTF2012}. The effect of operator ordering, appearing in the higher orders of MA, has been discussed in Refs.~\cite{QuesadaS2014,QuesadaS2015,KrummSV2016}.
	
The central idea of the present article is to apply the formalism of the Bloch-Messiah decomposition and the Magnus expansion to the simplest case of a broadband PDC, that of a type-I PDC with undepleted monochromatic plane-wave pump, for which an exact analytical solution is available. Using this analytical solution will allow us to investigate the effects of the higher-order terms of the Magnus expansion on the degree of squeezing and the parameters of the squeezing eigenmodes. We expect also that our results are qualitatively valid for the case of quasi-monochromatic pump and can serve as a limiting case for a broadband pump, where only numerical results are available.
				
	The article is organized as follows. In Sec.~\ref{Sec::Monochromatic_Pump_model} we describe the model of a type-I PDC with undepleted monochromatic plane-wave pump and its exact solution. In Sec.~\ref{Sec::BMR} we present a matrix formulation of this model, provide its Bloch-Messiah decomposition and define the squeezing eigenmodes as linear combinations of the monochromatic modes with opposite detunings from the central frequency. In terms of squeezing eigenmodes the state of the output field is a direct product of squeezed states for each mode, which is a great advantage of this particular choice of modal decomposition. In Sec.~\ref{Sec::Magnus_Expansion_Approximation} we apply the Magnus expansion to the quantum-mechanical evolution operator and obtain analytic expressions for the first three orders of MA. Then we compare these approximations with the exact solution, in particular with respect to the parameters, characterizing the degree of squeezing and the eigenmodes. In Sec.~\ref{Sec::Summary} we provide our conclusions and give some outlooks for the future.

\section{PDC with monochromatic pump}\label{Sec::Monochromatic_Pump_model}

We consider the process of collinear type-I PDC in a nonlinear $\chi^{(2)}$ crystal with a plane-wave monochromatic pump of frequency $\omega_p$. We shall assume that the pump wave is strong enough and is undepleted. A coordinate system is chosen with the $z$-axis in the direction of the pump wave propagation and with the origin at the front edge of the crystal. The pump is considered as a classical monochromatic wave $E^{(+)}_p(t,z) = E_pe^{i(k_pz-\omega_pt)}$, with the amplitude $E_p$, the wave vector $k_p$, and the frequency $\omega_p$. The down-converted wave is collinear with the pump wave but has a broadband spectrum of frequencies $\Omega$ around the central frequency $\omega_0=\omega_p/2$, with the corresponding wave vector $k_0$. The down-converted wave is described by the positive-frequency operator $E^{(+)}(t,z)$ normalized to photon-flux units, which can be decomposed into Fourier components as
\begin{align}
E^{(+)}(t,z) = \frac1{2\pi}\int a(\Omega,z)e^{i\left[k_0z-(\omega_0+\Omega)t\right]}d\Omega,
\end{align}
where $a(\Omega,z)$ is the photon annihilation operator with the frequency $\omega_0+\Omega$ and the longitudinal coordinate $z$. This operator describes the field at the frequency $\omega_0+\Omega$ as an operator-valued sideband component of the carrier wave at the frequency $\omega_0$. We shall call it below the sideband operator. It satisfies the canonical commutation relation
\begin{equation}\label{commutator}
[a(\Omega,z),a^\dagger(\Omega',z)] = 2\pi\delta(\Omega-\Omega'),
\end{equation}
where $a^\dagger(\Omega,z)$ is Hermitian conjugate of $a(\Omega,z)$.

We shall use another operator, $\epsilon(\Omega,z)$, defined by the relation~\cite{Kolobov1999}
\begin{equation}\label{aeps}
a(\Omega,z) = \epsilon(\Omega,z)e^{i\left(k(\Omega)-k_0\right)z},
\end{equation}
where $k(\Omega)$ is the wave vector of the down-converted light in the crystal corresponding to the frequency $\omega_0+\Omega$. Operator $\epsilon(\Omega,z)$ is convenient for the description of the nonlinear interaction inside the crystal and is a quantum-mechanical analog of the classical slowly-varying amplitude~\cite{Boyd}.
		
The evolution of the down-converted wave in the crystal is described by the equation~\cite{Boyd,Kolobov1999}
\begin{align}
		\partial_z \epsilon(\Omega,z)=\sigma e^{i\Delta(\Omega)z}
		 \epsilon^\dagger(-\Omega,z),
		 \label{Eq::PDC_DEq}
\end{align}
with the initial condition $\epsilon(\Omega,0)$. Here
\begin{align}
		\Delta(\Omega)
		=k_{p}-
		k(\Omega)-k(-\Omega)
\end{align}
is the phase-mismatch function and $\sigma$ is a coupling constant proportional to the pump-field amplitude and the nonlinear susceptibility on the crystal. Equation~(\ref{Eq::PDC_DEq}) describes a process of conversion of a pump photon with the frequency $\omega_p$ into signal and idler photons with opposite sidebands $\Omega$ and $-\Omega$.

The solution of Eq.~(\ref{Eq::PDC_DEq}) has a form of a Bogoliubov transformation~\cite{Kolobov1999}
\begin{equation}\label{Bogoliubov_epsilon}
\epsilon(\Omega,L) = A(\Omega)\epsilon(\Omega,0) + B(\Omega)\epsilon^\dagger(-\Omega,0),
\end{equation}
with the complex coefficients $A(\Omega)$ and $B(\Omega)$ given by
\begin{eqnarray}\nonumber
A(\Omega) &=& e^{i\Delta L/2} \left[\cosh \left(\Gamma L\right)-i\frac{\Delta}{2\Gamma}\sinh \left(\Gamma L\right)\right], \\\label{AB}
B(\Omega) &=& e^{i\Delta L/2}\frac{\sigma}{\Gamma}\sinh\left(\Gamma L\right),
\end{eqnarray}
where $\Gamma=\sqrt{|\sigma|^2-(\Delta/2)^2}$. At perfect phase-matching, where $\Delta(\Omega)=0$, and in the band of frequencies around this frequency $\Gamma$ is real. Outside this band $\Gamma$ is purely imaginary, and thus the hyperbolic functions in Eq.~\eqref{AB} become trigonometric.  Note, that the frequency detuning $\Omega$ enters Eq.~(\ref{AB}) only through $\Delta(\Omega)$ which is an even function. Therefore, the functions $A(\Omega)$ and $B(\Omega)$ are also even.

The sideband operator undergoes similar Bogoliubov transformation
\begin{equation}\label{Bogoliubov_a}
a(\Omega,L) = U(\Omega)a(\Omega,0) + V(\Omega)a^\dagger(-\Omega,0),
\end{equation}
where
\begin{eqnarray}\label{UV}
U(\Omega) &=& A(\Omega)e^{i\left(k(\Omega)-k_0\right)L}, \\\nonumber
V(\Omega) &=& B(\Omega)e^{i\left(k(\Omega)-k_0\right)L}.
\end{eqnarray}

The Bogoliubov transformation~(\ref{Bogoliubov_a}) is fully characterized by four real parameters. Indeed, Eq.~(\ref{Bogoliubov_a}) together with its Hermite conjugate with opposite detuning $-\Omega$ is described by four complex numbers $U(\pm\Omega)$, $V(\pm\Omega)$. Unitarity of Bogoliubov transformation imposes four real conditions $|U(\pm\Omega)|^2-|V(\pm\Omega)|^2=1$, and $U(\Omega)/V(\Omega)=U(-\Omega)/V(-\Omega)$ (the last complex equation provides two real conditions), so that only four real parameters remain. These four real parameters can be defined through the squeezing parameter, and three characteristic angles~\cite{Kolobov1999}
\begin{eqnarray}\label{r}
r(\Omega)& = &\ln\left(|U(\Omega)|+|V(\Omega)|\right),\\\label{psiL}
\psi_L(\Omega)& = &\frac12 \arg\left[U(\Omega)V(-\Omega)\right],\\\label{psi0}
\psi_0(\Omega)& = &\frac12 \arg\left[U^{-1}(\Omega)V(\Omega)\right],\\\label{kappa}
\kappa(\Omega)& = &\frac12 \arg\left[U(\Omega)U^{-1}(-\Omega)\right],
\end{eqnarray}
where the first three parameters are even functions of $\Omega$, while the fourth one is odd.
Below we explain the physical meaning of these four parameters. For each pair of modes with opposite detunings we construct~\cite{Kolobov1999} two input eigenquadrature operators
\begin{eqnarray}\label{quad1}
     X_1(\Omega,0)&=&a(\Omega,0)e^{-i\psi_0(\Omega)} +a^{\dagger}(-\Omega,0)e^{i\psi_0(\Omega)},\\
     \nonumber
     X_2(\Omega,0)&=&-i\left[a(\Omega,0)e^{-i\psi_0(\Omega)} -a^{\dagger}(-\Omega,0)e^{i\psi_0(\Omega)}\right],
\end{eqnarray}
and two output output eigenquadrature operators
\begin{eqnarray}\label{quad-out}
     X_1(\Omega,L)&=&a(\Omega,L)e^{-i\psi_L(\Omega)} +a^{\dagger}(-\Omega,L)e^{i\psi_L(\Omega)},\\
     \nonumber
     X_2(\Omega,L)&=&-i\left[a(\Omega,L)e^{-i\psi_L(\Omega)} -a^{\dagger}(-\Omega,L)e^{i\psi_L(\Omega)}\right].
\end{eqnarray}
In terms of these eigenquadratures the transformation~(\ref{Bogoliubov_a}) can be rewritten in a simple form
\begin{equation}\label{Xj}
     X_{j}(\Omega,L)=e^{\pm r(\Omega)+i\kappa(\Omega)}X_{j}(\Omega,0),
\end{equation}
where the plus (minus) sign corresponds to $j=1$ ($j=2$). It follows from Eq.~(\ref{Xj}) that the quadrature $X_2(\Omega,L)$ is squeezed below the standard quantum limit, while the conjugate quadrature $X_1(\Omega,L)$ is stretched above that limit. The squeezing parameter $r(\Omega)$ determines the degree of squeezing, while the angle of squeezing $\psi_L(\Omega)$ determines the choice of the coordinate axes on the complex plane for the eigenquadrature component at the output of the nonlinear crystal.

The angle $\psi_0(\Omega)$ determines the respective eigenquadrature component at the input to the nonlinear crystal. For the spontaneous PDC considered in this article this angle is irrelevant, since all quadratures of the input field are in the vacuum state. However, for the PDC with nonzero classical input or an input quantum state different from the vacuum this angle becomes important.

The last parameter $\kappa(\Omega)$ in our case of even $A(\Omega)$ and $B(\Omega)$ is independent of the nonlinear properties of the crystal and is given by
\begin{equation}\label{kappa2}
    \kappa(\Omega)=\frac{1}{2}\left[k(\Omega)-k(-\Omega)\right]L\approx \tau_g\Omega,
\end{equation}
where $\tau_g=L/v_g$ is the characteristic time during which the down-converted wave travels through the crystal at the group velocity $v_g=1/k'(0)$. Thus, the angle $\kappa(\Omega)$ describes the effect of the group delay due to crystal dispersion.

Substituting Eq.~(\ref{AB}) into Eqs.~(\ref{UV}), (\ref{psiL}) and (\ref{psi0}), and denoting $\varphi=\arg{\sigma}$, we obtain
\begin{eqnarray}\label{psipsi}
    \psi_L(\Omega) &=& \varphi-\psi_0(\Omega) \\\nonumber
    &=& \frac{\varphi}{2}+\frac12 \arg\left[\cosh \left(\Gamma L\right)-\frac{i\Delta}{2\Gamma}\sinh \left(\Gamma L\right)\right]\\\nonumber
    &+& \frac{1}{2} \arg\left[\frac{1}{\Gamma}\sinh \left(\Gamma L\right)\right].
\end{eqnarray}
This equation indicates that due to the symmetry of our system, two angles $\psi_L(\Omega)$ and $\psi_0(\Omega)$ are not independent. Therefore, in what follows we shall provide the results only for the angle $\psi_L(\Omega)$ at the output of the crystal. It is worth noting that this symmetry manifests itself due to particular choice of our PDC scheme, and is not necessarily present in all PDC processes. For example, for PDC in aperiodically poled quasi-phase-matched crystals this additional symmetry is lifted, and the angles $\psi_L(\Omega)$ and $\psi_0(\Omega)$ become independent~\cite{Horoshko2013,Horoshko2017}.

\begin{figure}[ht]
		\includegraphics[width=8.6cm]{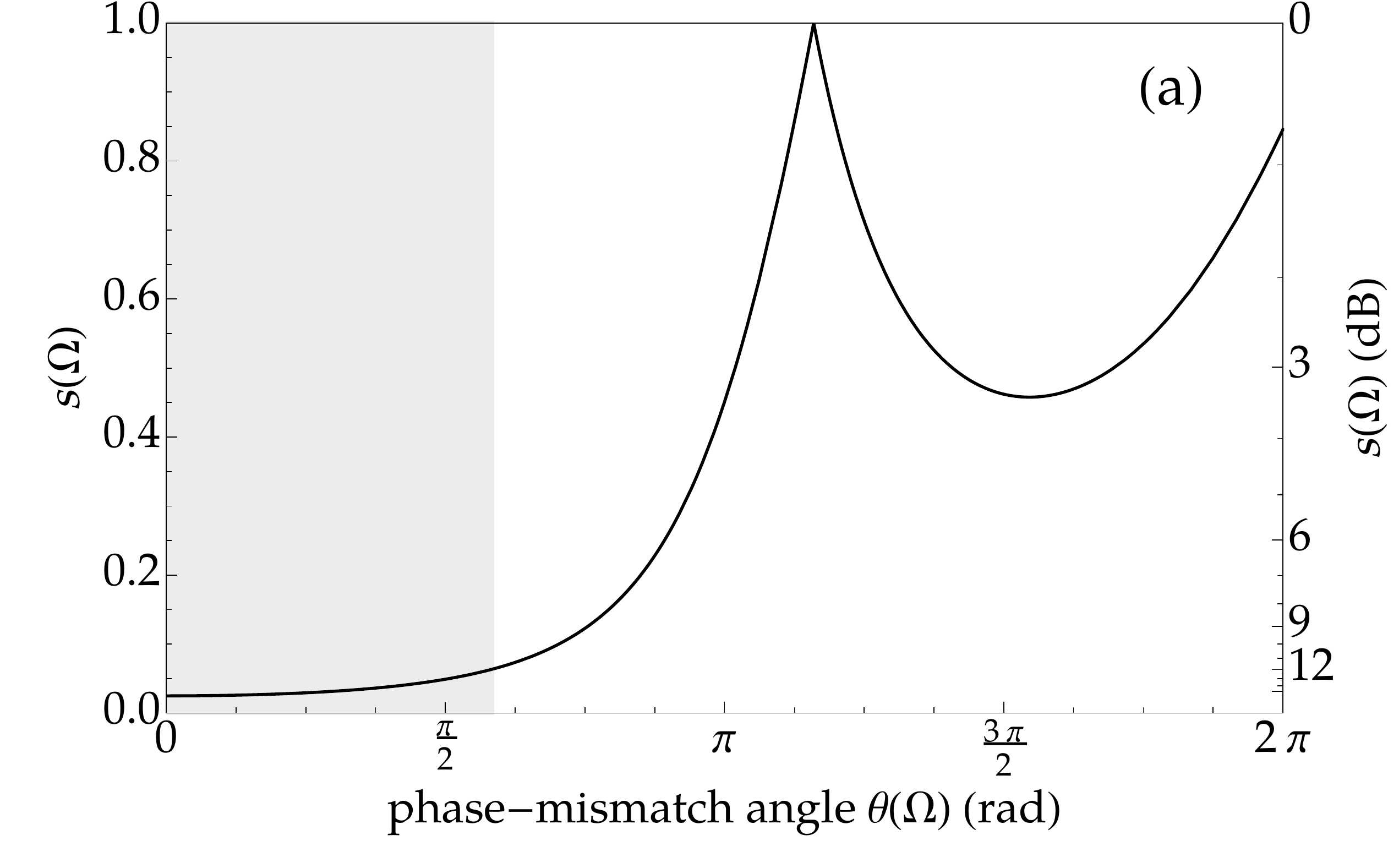}\\
		\includegraphics[width=8.6cm]{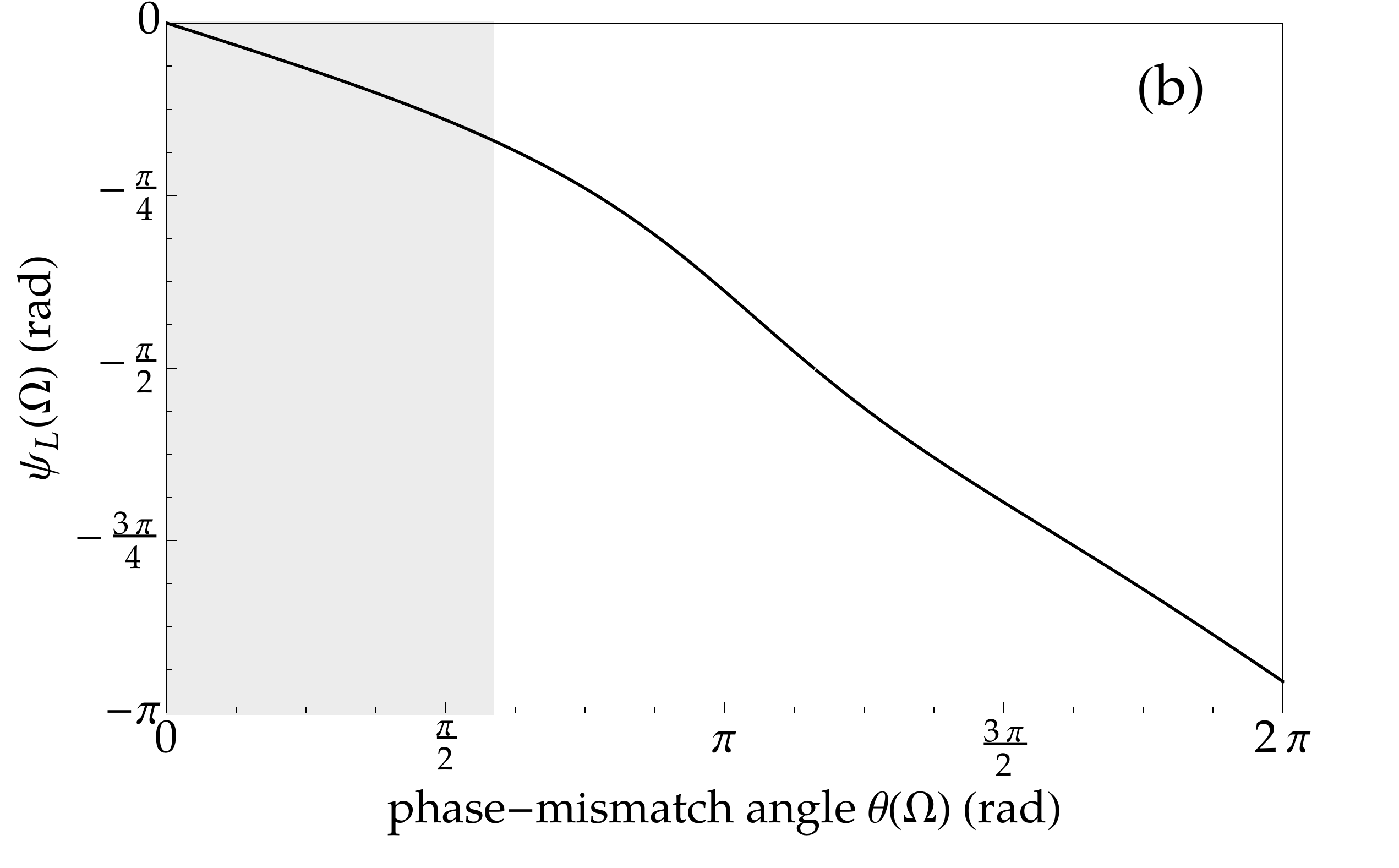}\\
		\caption{Graphs of (a) the spectrum of squeezing $s(\Omega)$  and (b) the angle of squeezing $\psi_L(\Omega)$ as  functions of the phase-mismatch angle for the exact solution. The gain exponent is $g=1.84$, corresponding to $16 \, \mathrm{dB}$ of maximal squeezing. The phase of the pump is chosen so that $\varphi=0$.  The gray area indicates the band, where $\Gamma$ is real.
			}\label{Fig::psiL_r_exact}
   \end{figure}

The correlation function of the squeezed quadrature components $X_{j}(\Omega,L)$ at the output of the crystal can be calculated from that at its input using Eq.~(\ref{Xj}). For the vacuum field at the input we have for both quadratures
\begin{equation}\label{corrin}
      \langle X_{j}(\Omega,0)X_{j}(\Omega',0)\rangle=2\pi \delta(\Omega+\Omega'),
\end{equation}
and therefore at the output
\begin{eqnarray}\label{corr}
      \langle X_{1}(\Omega,L)X_{1}(\Omega',L)\rangle &=& \frac{2\pi}{s(\Omega)} \delta(\Omega+\Omega'),\\\nonumber
      \langle X_{2}(\Omega,L)X_{2}(\Omega',L)\rangle &=& 2\pi s(\Omega)\delta(\Omega+\Omega'),
\end{eqnarray}
where $s(\Omega)=\exp[-2r(\Omega)]$ is known as the spectrum of squeezing. The spectrum of squeezing together with the angle of squeezing are shown in Fig.~\ref{Fig::psiL_r_exact} as functions of the phase-mismatch angle $\theta(\Omega)=\Delta(\Omega)L/2$. The amplitude of the pump is characterized by the parametric gain exponent $g=|\sigma|L$.

The angle in Fig.~\ref{Fig::psiL_r_exact}b and in the subsequent figures of Sec.~\ref{Sec::Magnus_Expansion_Approximation} is the continuous version of the angle of squeezing $\psi_L(\Omega)$. The original angle of squeezing $\psi_L(\Omega)$ in Eq.~(\ref{psiL}) experiences a jump of $\pi/2$ at the frequencies $\Omega$ where $r(\Omega)=0$. In Fig.~\ref{Fig::psiL_r_exact}b we have corrected for this jump in order to make the $\psi_L(\Omega)$ a continuous function. In other words, the continuous version of $\psi_L(\Omega)$ corresponds to the stretched quadrature between the odd and the even zeros of $r(\Omega)$.

From Fig.~\ref{Fig::psiL_r_exact}a we see that squeezing is maximal for perfect phase matching when the phase-mismatch angle is zero, $\theta(\Omega)=0$. For increasing mismatch it shows oscillations, decreasing in magnitude until disappearing completely for very large values of $\theta(\Omega)$ (not shown in Fig.~\ref{Fig::psiL_r_exact}a). The angle of squeezing decreases monotonically with $\theta(\Omega)$ approaching its asymptotic value $\psi_L(\Omega)\to-\theta(\Omega)/2$.

\section{Bloch-Messiah decomposition}\label{Sec::BMR}
\subsection{Matrix formulation}

The theory of PDC developed in the previous section can be formulated in a compact matrix form which will allow us to perform the Bloch-Messiah decomposition and the Magnus expansion. We collect the slowly-varying operators $\epsilon (\Omega)$ and $\epsilon (-\Omega)$ in a column vector as
	\begin{align}
		{\boldsymbol \xi}(z)=\left(\begin{array}{c}  \epsilon (\Omega,z)\\  \epsilon (-\Omega,z)\\ \epsilon^\dagger (\Omega,z)\\  \epsilon^\dagger (-\Omega,z)\end{array}\right),
	\end{align}
and write Eq.~\eqref{Eq::PDC_DEq} in a matrix form as
 	\begin{align}
		\partial_z\boldsymbol{\xi}(z)=
		 -i\mathbf{F}(z)\boldsymbol{\xi}(z)
		 \text{.}
		 \label{Eq::PDC_linear_matrix_DEq}
	\end{align}
Here and below, when it does not create ambiguity, we shall omit the arguments $\Omega$ and $-\Omega$ in order to simplify the notations.
The coupling matrix $\mathbf{F}$ is given by
	\begin{align}
		\mathbf{F}(z)= \begin{pmatrix}
				0					&	i\sigma e^{i\Delta z}P	\\
				i\sigma^* 	e^{-i\Delta z}P		&	0
		           \end{pmatrix}
		\text{,}
		\label{Eq::Coupling_Matrix_F}
	\end{align}
where
	\begin{align}
		P=
		\begin{pmatrix}
			0	&	1	\\
			1	&	0	
		\end{pmatrix}
		\text{.}
	\end{align}

The operators ${\boldsymbol \xi}^\text{out}$ at the output of the crystal, $z=L$, are related with the operators ${\boldsymbol \xi}^\text{in}$ at its input, $z=0$, by a linear matrix transformation ${\boldsymbol \xi}^\text{out}=\mathbf{S}{\boldsymbol \xi}^\text{in}$  with the matrix $\mathbf{S}$ given by
	\begin{align}
		\mathbf{S}=&\begin{pmatrix}
				A(\Omega) I	&	B(\Omega) P	\\
				B(\Omega)^* P	&	A(\Omega)^*I	
		           \end{pmatrix}
		\text{.}
		\label{Eq::Linear_Matrix_Transform_Input_to_Output}
	\end{align}
This linear transformation preserves the commutation relations of the operators ${\boldsymbol \xi}^\text{in}$, and therefore matrix $\mathbf{S}$ is a symplectic matrix~\cite{ArvindDM1995}, satisfying the relation $\mathbf{S}\mathbf{K}\mathbf{S}^\dagger=\mathbf{K}$, with
	\begin{align}
		\mathbf{K}=
		\begin{pmatrix}
			I	&	0	\\
			0	&	-I
		\end{pmatrix}
		\text{.}
		\label{Eq::Symplectic_Condition}
	\end{align}
	
Similarly, we introduce a column vector for the sideband operators $a(\Omega,z)$ and $a(-\Omega,z)$ as
	\begin{align}
		{\boldsymbol a}(z)=\left(\begin{array}{c}  a (\Omega,z)\\  a (-\Omega,z)\\ a^\dagger (\Omega,z)\\  a^\dagger (-\Omega,z)\end{array}\right).
	\end{align}
This vector, due to Eq.~(\ref{aeps}), can be written as ${\boldsymbol a}(z)=\boldsymbol\Phi_z{\boldsymbol \xi}(z)$, where the unitary matrix $\boldsymbol\Phi_z$ is defined as
	\begin{equation}\label{Phiz}
	  \boldsymbol\Phi_z = \diag\{e^{i\delta k(\Omega)z},e^{i\delta k(-\Omega)z},e^{-i\delta k(\Omega)z},e^{-i\delta k(-\Omega)z}\},
	\end{equation}
with $\delta k(\Omega)=k(\Omega)-k_0$. In terms of the sideband operators the exact solution is written as ${\boldsymbol a}^\text{out}=\mathbf{\tilde S}{\boldsymbol a}^\text{in}$, where the complex symplectic matrix $\mathbf{\tilde S}=\boldsymbol\Phi_L\mathbf{S}$ is expressed through the four real parameters given by Eqs.~(\ref{r})-(\ref{kappa}) as
	\begin{align}
		\mathbf{\tilde S}=&\begin{pmatrix}
				e^{i(\psi_L-\psi_0)}\cosh(r)\Lambda	    &	e^{i(\psi_L+\psi_0)}\sinh(r)\Lambda P	\\
				e^{-i(\psi_L+\psi_0)}\sinh(r)\Lambda^* P	&	e^{i(\psi_0-\psi_L)}\cosh(r)\Lambda^*	
		           \end{pmatrix}
		\text{,}
		\label{Stilde}
	\end{align}
where
    \begin{align}
		\Lambda=&\begin{pmatrix}
				e^{i\kappa}    &	0	\\
				0	&	e^{-i\kappa}	
		           \end{pmatrix}
		\text{.}
		\label{Lambda}
	\end{align}

Equation~\eqref{Stilde} is the complex symplectic representation of the Bogoliubov transformation Eq.~\eqref{Bogoliubov_a}.

\subsection{Bloch-Messiah decomposition and the squeezing eigenmodes}

Bloch-Messiah decomposition in our case consists in factorization of the symplectic matrix $\mathbf{\tilde S}$ in a product of three matrices \cite{Braunstein2005}
	\begin{align}
		\mathbf{\tilde S}=\mathbf{V}\mathbf{D}(r)\mathbf{W}^\dagger
		\text{,}
		\label{Eq::Bloch_Messiah_decomposition}
	\end{align}
where the unitary $4\times4$ matrices $\mathbf{V}$ and $\mathbf{W}$ have the structure
	\begin{align}
		\mathbf{V}&=	\begin{pmatrix}
						V	&	0	\\
						0	&	V^*
					\end{pmatrix}
					\text{, }
		\mathbf{W}=	\begin{pmatrix}
						W	&	0	\\
						0	&	W^*
					\end{pmatrix}
					\text{, }
	\end{align}
and the real $4\times4$ matrix $\mathbf{D}(r)$ is given by
    \begin{align}
		\mathbf{D}(r)&=\begin{pmatrix}
						\cosh\left(r\right)I	&	\sinh\left(r\right)I	\\
						\sinh\left(r\right)I	&	\cosh\left(r\right)I
				\end{pmatrix}
		=\exp
					 \begin{pmatrix}
						0	&	r I	\\
						r I	&	0
					\end{pmatrix}
		\text{.}
		\label{Eq::decomposition_matrices}
	\end{align}
	The $2\times2$ matrices $V$ and $W$ are defined as follows
	\begin{align}
		V&=	\frac{e^{i\psi_L}}{\sqrt{2}}
					\begin{pmatrix}
						e^{i\kappa}	&	0	\\
						0	&	e^{-i\kappa}
					\end{pmatrix}
					\begin{pmatrix}
						1	&	i	\\
						1	&	-i
					\end{pmatrix}
					\text{, }
		W&=	\frac{e^{i\psi_0}}{\sqrt{2}}
					\begin{pmatrix}
						1	&	i	\\
						1	&	-i
					\end{pmatrix}
		\text{,}
		\label{Eq::VW}
	\end{align}
where the three characteristic angles are taken at the detuning $\Omega$.
	
Bloch-Messiah decomposition allows us to define the squeezing eigenmodes for the output field by the relation
\begin{eqnarray}
&&\left(\begin{array}{c} b_c(\Omega,L) \\ b_s(\Omega,L) \end{array} \right) = V^\dagger \left(\begin{array}{c} a(\Omega,L) \\ a(-\Omega,L) \end{array} \right)\\\nonumber
&&= \frac{e^{-i\psi_L}}{\sqrt{2}}\left(\begin{array}{c} a(\Omega,L)e^{-i\kappa} + a(-\Omega,L)e^{i\kappa} \\ -ia(\Omega,L)e^{-i\kappa} + ia(-\Omega,L)e^{i\kappa} \end{array} \right).
\end{eqnarray}

According to Eq.~(\ref{Eq::Bloch_Messiah_decomposition}) the annihilation operators of these modes are expressed as
\begin{eqnarray}\label{b1b2}
b_c(\Omega,L) &=& \cosh(r)b_{c}(\Omega,0) + \sinh(r)b^\dagger_{c}(\Omega,0), \\\nonumber
b_s(\Omega,L) &=& \cosh(r)b_{s}(\Omega,0) + \sinh(r)b^\dagger_{s}(\Omega,0),
\end{eqnarray}
via the input vacuum modes with the operators $b_{c}(\Omega,0)$ and $b_{s}(\Omega,0)$, defined as
\begin{eqnarray}
\left(\begin{array}{c}b_{c}(\Omega,0) \\ b_{s}(\Omega,0) \end{array} \right) &=& W^\dagger \left(\begin{array}{c} a(\Omega,0) \\ a(-\Omega,0) \end{array} \right)\\\nonumber
&=& \frac{e^{-i\psi_0}}{\sqrt{2}}\left(\begin{array}{c} a(\Omega,0) + a(-\Omega,0) \\ -ia(\Omega,0) + ia(-\Omega,0) \end{array} \right).
\end{eqnarray} 

As follows from Eq.~(\ref{b1b2}), the eigenmodes described by operators $b_c(\Omega,L)$ and $b_s(\Omega,L)$ are squeezed along the same direction in the phase space with the same degree of squeezing $r(\Omega)$.

We stress the difference between the modes described by the operators $a(\Omega,L)$ and $a(-\Omega,L)$ from the eigenmodes described by the operators $b_c(\Omega,L)$ and $b_s(\Omega,L)$: the first ones are in a \textit{two-mode squeezed state} and, therefore, entangled, while the second are in a \textit{single-mode squeezed state} each and therefore statistically independent. This is the reason why we call these modes as {\it squeezing eigenmodes}.

The modal functions corresponding to the squeezing eigenmodes, $f_c(t,z|\Omega)$ and $f_s(t,z|\Omega)$, are given by 
\begin{eqnarray}\label{ff}
&&\left(\begin{array}{c} f_c(t,z|\Omega) \\ f_s(t,z|\Omega) \end{array} \right) = V^\dagger \left(\begin{array}{c} e^{i\Omega t} \\ e^{-i\Omega t} \end{array} \right) \frac{e^{-i\left(k_0z-\omega_0t\right)}}{2\pi}\\\nonumber
&&= \frac{\sqrt{2}}{2\pi}e^{-i\left(k_0z-\omega_0t+\psi_L\right)}\left(\begin{array}{c} \cos\left(\Omega t -\kappa(\Omega)\right) \\ \sin\left(\Omega t -\kappa(\Omega)\right) \end{array} \right).
\end{eqnarray}
The spectral profiles of these modes include two delta-functions at the frequencies $\omega_0-\Omega$ and $\omega_0+\Omega$, thus the squeezing eigenmodes are bichromatic. We note that the modal functions, Eq.~\eqref{ff}, are the functions of time $t$ and the longitudinal coordinate $z$, while the frequency $\Omega$ and the indices $c,s$ serve as the mode markers, equivalent to an integer index in the case of discrete modes. The modal functions, Eq.~\eqref{ff}, are even or odd functions of $\Omega$. Therefore the frequency $\Omega$ in the eigenmode definition is restricted to the non-negative values only. The modal functions with negative frequencies are linearly dependent on the positive-frequency ones. It means that the corresponding eigenmodes are redundant in the modal decomposition of the field and can be omitted.

Now we can introduce the Hermitian operators for the generalized ``position'' and ``momentum'' of the squeezing eigenmodes
\begin{eqnarray}\label{qp}
q_{c,s}(\Omega,z) &=& \frac{b_{c,s}(\Omega,z) + b_{c,s}^\dagger(\Omega,z)}{\sqrt{2}},\\\nonumber
p_{c,s}(\Omega,z) &=& \frac{-ib_{c,s}(\Omega,z) + ib_{c,s}^\dagger(\Omega,z)}{\sqrt{2}},
\end{eqnarray}
where $z$ in the framework of current discussion takes only values $0$ and $L$. Using these Hermitian operators we can write the non-Hermitian quadrature operators in Eq.~\eqref{quad1} as
\begin{eqnarray}\label{XviaQP}
X_1(\Omega,0) &=& q_c(\Omega,0) + iq_s(\Omega,0),\\\nonumber
X_2(\Omega,0) &=& p_c(\Omega,0) + ip_s(\Omega,0),\\\nonumber
X_1(\Omega,L) &=& \left[q_c(\Omega,L) + iq_s(\Omega,L)\right]e^{i\kappa},\\\nonumber
X_2(\Omega,L) &=& \left[p_c(\Omega,L) + ip_s(\Omega,L)\right]e^{i\kappa}.
\end{eqnarray}
We see that the quadrature operator $X_1(\Omega,L)$ combines the position operators of two squeezing eigenmodes, while the quadrature operator $X_2(\Omega,L)$ combines their momentum operators. This explains why the quadratures $X_1(\Omega,L)$ and $X_2(\Omega,L)$ are complementary, and cannot be measured simultaneously. The transformation of the operators, defined by Eq.~\eqref{qp} in the nonlinear crystal corresponds to single-mode squeezing:
\begin{eqnarray}\label{qp-out}
q_{c,s}(\Omega,L) &=& e^rq_{c,s}(\Omega,0),\\\nonumber
p_{c,s}(\Omega,L) &=& e^{-r}p_{c,s}(\Omega,0),
\end{eqnarray}
and can be interpreted as modulation of quantum fluctuations in the nonlinear interaction~\cite{Kolobov1999}.

In conclusion, in this section we have formulated the Bloch-Messiah decomposition of the Bogoliubov transformation. Using this decomposition, we have defined the squeezing eigenmodes and demonstrated that the parameters $\kappa(\Omega)$ and $\psi_L(\Omega)$ define the modal functions, while the squeezing parameter $r(\Omega)$ determines the degree of squeezing. The last parameter $\psi_0(\Omega)$ defines the input vacuum modes. In the next section we shall explore the behavior of these parameters in different orders of the Magnus expansion.

\section{Magnus approximation}~\label{Sec::Magnus_Expansion_Approximation}

\subsection{Definition of the Magnus expansion}
	
The solution of Eq.~\eqref{Eq::PDC_linear_matrix_DEq} can be formally written in the form of a $\mathcal{T}$-exponent~\cite{VogelW2006}
	\begin{align}
		\mathbf{S}=\mathcal{T}e^{-i\int_0^L dz\, \mathbf{F}(z)}
		\text{,}
		\label{Eq::T_exponent_definition}
	\end{align}
where the symbol $\mathcal{T}$ denotes a $z$-ordering operator,
putting the operators with higher $z$-values to the left in the expansion of the exponent.
	
Decomposing $\ln\mathbf{S}$ in the Taylor series in the modulus of the coupling constant $|\sigma|$, one can represent the $\mathcal{T}$-exponent in the form of the Magnus expansion~\cite{BlanesaCOR2009}
	\begin{align}
		\mathbf{S}=e^{ \boldsymbol\Omega_1 + \boldsymbol\Omega_2 + \boldsymbol\Omega_3 + \dots}
		\text{,}
		\label{Eq::Magnus_Expansion}
	\end{align}
where $\boldsymbol\Omega_k$ is a $4\times4$ matrix proportional to $|\sigma|^k$. The first three terms in Eq.~(\ref{Eq::Magnus_Expansion}) are
	\begin{eqnarray}\label{Omega1}
	 \boldsymbol\Omega_1 &=& -i\int_0^L dz\, \mathbf{F}(z), \\\label{Omega2}
	 \boldsymbol\Omega_2 &=& -\frac12\int_0^L dz_1\int_0^{z_1} dz_2\, [\mathbf{F}(z_1),\mathbf{F}(z_2)], \\\label{Omega3}
	 \boldsymbol\Omega_3 &=& \frac{i}6\int_0^L dz_1\int_0^{z_1} dz_2\int_0^{z_2} dz_3\, \\\nonumber &\times&\left([\mathbf{F}(z_1),[\mathbf{F}(z_2),\mathbf{F}(z_3)]]+[\mathbf{F}(z_3),[\mathbf{F}(z_2),\mathbf{F}(z_1)]]\right).
	\end{eqnarray}
Keeping the first $k$ terms in the Magnus expansion given by Eq.~(\ref{Eq::Magnus_Expansion}), we shall obtain an approximation of the $\mathcal{T}$-exponent in Eq.~(\ref{Eq::T_exponent_definition}) which we shall call the Magnus approximation (MA) of the $k$-th order,
	\begin{align}
		\mathbf{S}_k=\exp\left\{ \sum_{i=1}^k\boldsymbol\Omega_i\right\}
		\text{.}
		\label{Eq::Magnus_Expansion Approx}
	\end{align}

A remarkable property of this approximation is the symplectic structure of the approximate transformation matrix $\mathbf{S}_k$ for any $k$. This property of $\mathbf{S}_k$ implies conservation of the commutation relations for the creation and annihilation operators of the field for each order $k$. This feature of the MA represents a great advantage as compared to other approximate methods such as, for example, the Dyson expansion. In particular, it will guarantee that for the vacuum input state of PDC the output state for each order $k$ of the MA will be a squeezed state with the respective four real parameters defined above.

Therefore, for each order $k$ of the MA we shall define a respective symplectic matrix $\mathbf{\tilde S}_k = \boldsymbol\Phi_L\mathbf{S}_k$, for the transformation of the sideband operators, which can be parameterized by four real parameters $\{r_k(\Omega),\psi_{L,k}(\Omega),\psi_{0,k}(\Omega),\kappa_{k}(\Omega)\}$, similarly to the parametrization of the exact solution in Eq.~\eqref{Stilde}.	

\subsection{First-order Magnus approximation}

The first-order MA is obtained by keeping only the term $\boldsymbol\Omega_1$ in Eq.~(\ref{Eq::Magnus_Expansion}), which is equivalent to neglecting the $z$-ordering in Eq.~(\ref{Eq::T_exponent_definition}). Substituting Eq.~\eqref{Eq::Coupling_Matrix_F} into Eq.~\eqref{Omega1}, and performing the integration we obtain
	\begin{align}\label{Omega1int}
		\boldsymbol\Omega_1 = \left(\begin{array}{cc} 0 & b_1e^{i(\varphi+\theta)}P \\ b_1e^{-i(\varphi+\theta)}P & 0 \end{array} \right),
	\end{align}
where $b_1=g\sinc\theta$. Calculating the exponent of Eq.~\eqref{Omega1int} as power series and summing up even and odd powers separately, we arrive at
\begin{equation}
\mathbf{S}_1 = e^{\boldsymbol\Omega_1} = \left(\begin{array}{cc} I\cosh{b_1} & Pe^{i(\varphi+\theta)}\sinh{b_1} \\ Pe^{-i(\varphi+\theta)}\sinh{b_1} & I\cosh{b_1} \end{array} \right).
\end{equation}

Symplectic matrix $\mathbf{S}_1$ determines the transformation of the slowly-varying amplitudes $\boldsymbol\xi(z)$. Passing to the symplectic matrix $\mathbf{\tilde S}_1$, for the sideband operators $\mathbf{a}(z)$, we have
\begin{equation}\label{Stilde1}
\mathbf{\tilde S}_1 =  \left(\begin{array}{cc} e^{-i\theta}\Lambda\cosh{b_1} & \Lambda Pe^{i\varphi}\sinh{b_1} \\ \Lambda^*Pe^{-i\varphi}\sinh{b_1} & e^{i\theta}\Lambda^*\cosh{b_1} \end{array} \right).
\end{equation}

Comparing Eq.~\eqref{Stilde1} with Eq.~\eqref{Stilde} we conclude that in the first-order MA the parameter $\kappa(\Omega)$ is the same as in the exact solution, and the relation $\psi_{0,1}(\Omega) = \varphi - \psi_{L,1}(\Omega)$ holds, as well. As for the other two parameters, characterizing the Bogoliubov transformation, they are different:
    \begin{eqnarray}\label{r1}
    r_1(\Omega) &=& g|\sinc\theta|, \\
    \psi_{L,1}(\Omega) &=& \frac12(\varphi-\theta)+\frac{1}{2}\arg(\sinc\theta).\notag
    \end{eqnarray}
We remind that the phase-mismatch angle $\theta(\Omega)$ is a function of the frequency $\Omega$, as defined in Sec.~\ref{Sec::Monochromatic_Pump_model}.

For the frequencies $\Omega$ of the perfect phase matching, where $\Delta(\Omega)=0$, one can easily find $r_1=g=r$ and $\psi_{L,1} = \psi_{L}$. Similarly, for the frequencies $\Omega$ where $\Delta(\Omega)\rightarrow\pm\infty$, we find $r,r_1\to0$, $\psi_{L} \to \psi_{L,1}$. For other frequencies these parameters are, in general, different.
	
\subsection{Second-order Magnus approximation}

In the second-order MA we keep the two first terms $\boldsymbol\Omega_1$ and $\boldsymbol\Omega_2$ in Eq.~(\ref{Eq::Magnus_Expansion}). For calculating the second term we evaluate the commutator
\begin{equation}
[\mathbf{F}(z_1),\mathbf{F}(z_2)] = 2i|\sigma|^2\sin\left(\Delta(\Omega)(z_2-z_1)\right)\mathbf{K},
\end{equation}
and, integrating it according to Eq.~\eqref{Omega2}, we obtain
\begin{align}\label{Omega2int}
\boldsymbol\Omega_2 = \frac{ig^2}{2}\left[ j_0(\theta)\sin\theta- j_1(\theta)\cos\theta\right] \mathbf{K}.
\end{align}
Here $j_m(\theta)$ is the spherical Bessel function, i.~e.~, $j_0(\theta)=\sinc\theta$, $j_1(\theta)=(\sinc\theta-\cos\theta)/\theta$, etc.

Evaluating the exponent of $\boldsymbol\Omega_1+\boldsymbol\Omega_2$ and multiplying the result by $\boldsymbol\Phi_L$ we obtain the second-order approximation of the symplectic matrix
\begin{equation}\label{Stilde2}
\mathbf{\tilde S}_2 = \left(\begin{array}{cc} U_2 & V_2 \\
V_2^* & U_2^* \end{array} \right),
\end{equation}
with the $2\times2$ matrices $U_2$ and $V_2$ defined as
\begin{eqnarray}\label{UV2}
U_2 &=& \Lambda e^{-i\theta}\left(\cosh{\gamma_2}+\frac{ia_2}{\gamma_2}\sinh{\gamma_2}\right),\\\nonumber
V_2 &=& \Lambda Pe^{i\varphi} \frac{b_2}{\gamma_2}\sinh{\gamma_2},
\end{eqnarray}
and
	\begin{align}\label{ab2}
		a_2&=\frac{g^2}{2}\left[ j_0(\theta)\sin\theta- j_1(\theta)\cos\theta\right],\\\notag
		b_2&=gj_0(\theta),\\\notag
        \gamma_2&=\sqrt{b_2^2-a_2^2}
		\text{.}
	\end{align}

We observe from Eq.~\eqref{UV2} that the phase of $V_2$ is equal to $\varphi$ and that $\gamma_2$ may become imaginary if $|a_2|>|b_2|$, but $\sinh\gamma_2/\gamma_2$ is always real. Therefore, the relation $\psi_{0,2}(\Omega)=\varphi-\psi_{L,2}(\Omega)$ holds in the second-order approximation as well.

Comparing Eq.~\eqref{Stilde2} with Eq.~\eqref{Stilde} we also conclude that the parameter $\kappa$ in the second-order approximation is that of the exact solution, $\kappa_2(\Omega)=\kappa(\Omega)$.

The two remaining parameters are
\begin{eqnarray}\nonumber
r_2(\Omega) &=& \ln\left\{ \left|\cosh{\gamma_2}+\frac{ia_2}{\gamma_2}\sinh{\gamma_2}\right| + \left|\frac{b_2}{\gamma_2}\sinh{\gamma_2}\right|\right\},\\\label{rpsi2}
\psi_{L,2}(\Omega) &=& \frac\varphi2-\frac{\theta}{2} + \frac12\arg\left\{\cosh(\gamma_2)+\frac{ia_2}{\gamma_2}\sinh{\gamma_2}\right\}\notag\\
&+&\frac{1}{2}\arg(b_2)+\frac{1}{2}\arg\left(\frac{\sinh(\gamma_2)}{\gamma_2}\right)\text{.}
\end{eqnarray}
As in the first-order approximation, these parameters coincide with those of the exact solution at $\Delta(\Omega)=0$ and $\Delta(\Omega)\to\pm\infty$.

\subsection{Third-order Magnus approximation}

In the third-order Magnus approximation we keep the three first terms $\boldsymbol\Omega_1$, $\boldsymbol\Omega_2$ and $\boldsymbol\Omega_3$ in Eq.~(\ref{Eq::Magnus_Expansion}). After evaluating the corresponding commutators and performing the integration, we obtain
\begin{align}\label{Omega3int}
\boldsymbol\Omega_3 =\frac{g^3}{6}\left[j_0(\theta)+j_2(\theta)-j_0^3(\theta)\right]
\left(\begin{array}{cc} 0 & e^{i(\varphi+\theta)} P \\ e^{-i(\varphi+\theta)} P & 0 \end{array} \right) ,
\end{align}
and
\begin{equation}\label{Stilde3}
\mathbf{\tilde S}_3 = \left(\begin{array}{cc} U_3 & V_3 \\
V_3^* & U_3^* \end{array} \right)
\end{equation}
with the $2\times2$ matrices $U_3$ and $V_3$ defined as
\begin{eqnarray}\label{UV3}
U_3 &=& \Lambda e^{-i\theta}\left(\cosh{\gamma_3}+\frac{ia_3}{\gamma_3}\sinh{\gamma_3}\right),\\\nonumber
V_3 &=& \Lambda Pe^{i\varphi} \frac{b_3}{\gamma_3}\sinh{\gamma_3},
\end{eqnarray}
where
	\begin{align}\label{ab3}
		a_3&=\frac{g^2}{2}\left[ j_0(\theta)\sin\theta- j_1(\theta)\cos\theta\right],
	\\\notag
		b_3&=gj_0(\theta)+\frac{g^3}{6}\left[j_0(\theta)+j_2(\theta)-j_0^3(\theta)\right],\\\notag
    \gamma_3&=\sqrt{b_3^2-a_3^2}.
	\end{align}

We observe that the relations $\psi_{0,3}(\Omega)=\varphi-\psi_{L,3}(\Omega)$, $\kappa_3(\Omega)=\kappa(\Omega)$ hold in the third-order approximation as well, and that the other two parameters are given by equations, similar to Eq.~\eqref{rpsi2}.

We remind that the corrections of the MA higher than the first-order are due to non-zero commutators of the matrix $\mathbf{F}(z)$ with itself at different points $z$. Thus, deviations from the first-order MA are the manifestations of this non-commutativity, known in the literature as the operator ordering effects. Such effects were studied, for example, in  Refs.~~\cite{QuesadaS2014,QuesadaS2015,KrummSV2016}.

\subsection{Comparison of the three approximations}\label{Sec::Analysis}

\begin{figure}[ht]
		\includegraphics[width=8.6cm]{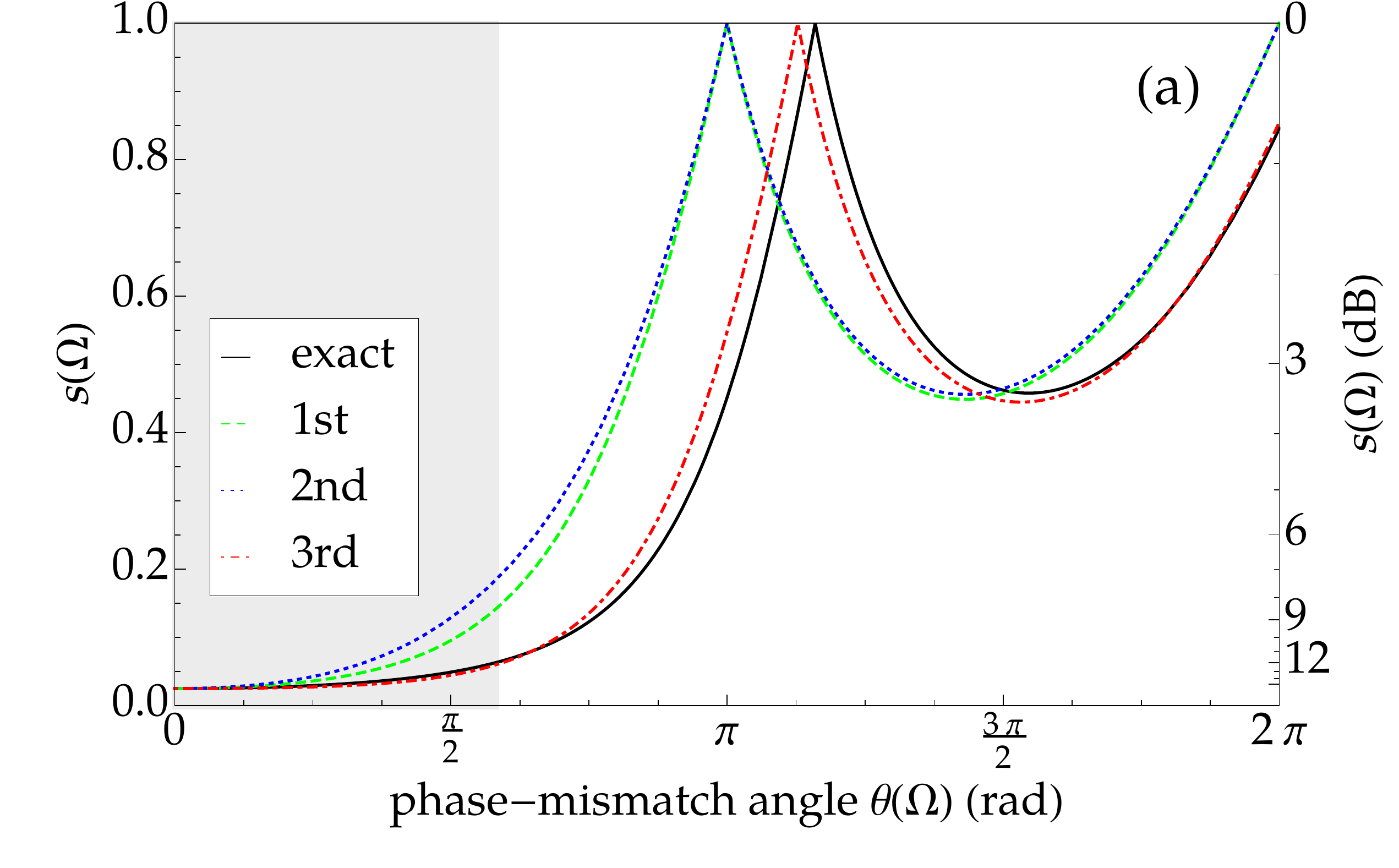}\\
		\includegraphics[width=8.6cm]{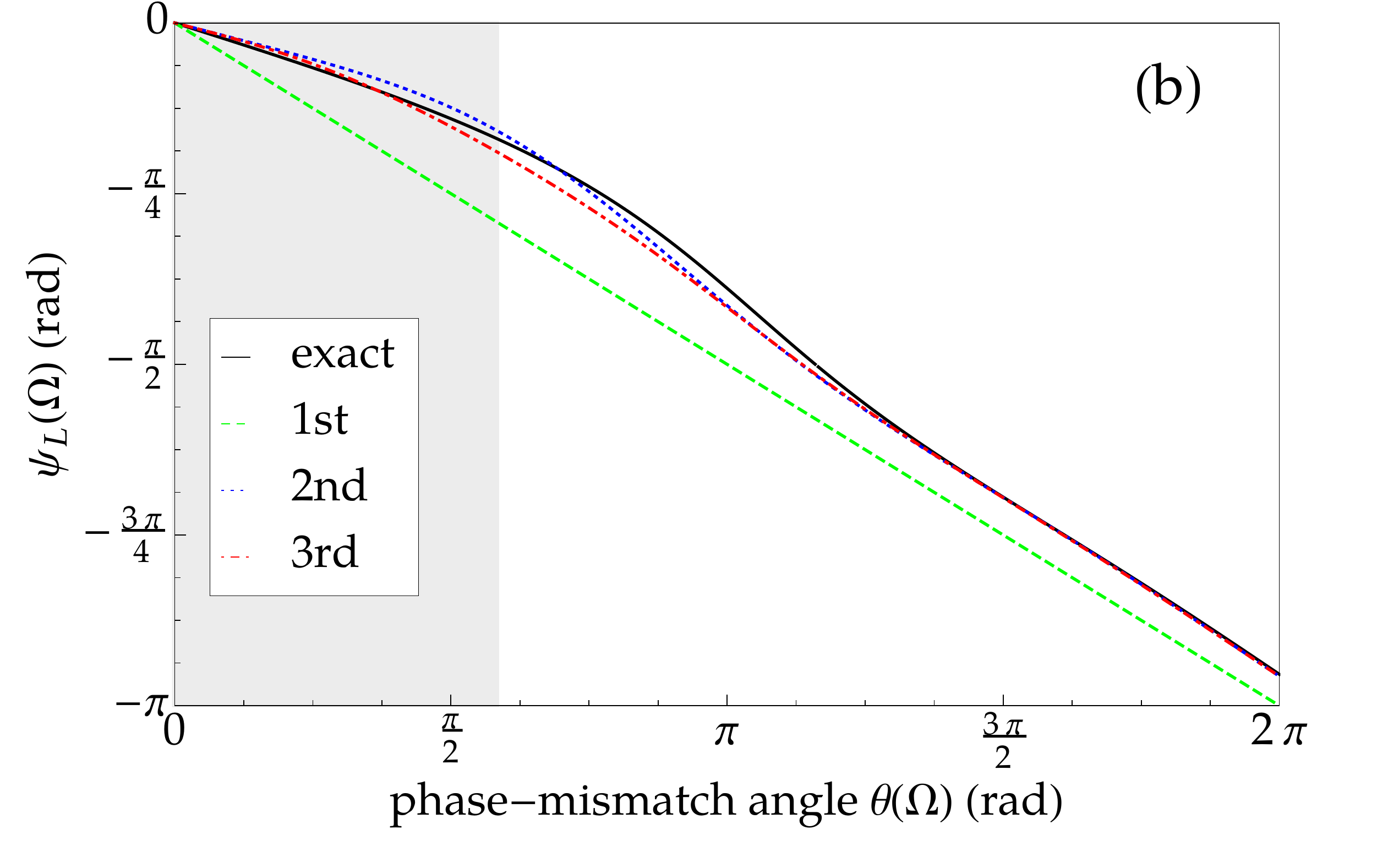}\\
		\caption{
			Graphs of (a) the spectrum of squeezing $s(\Omega)$  and (b) the angle of squeezing $\psi_L(\Omega)$ as functions of the phase-mismatch angle for the exact solution and the first three orders of the Magnus approximation. The parametric gain exponent is $g=1.84$, corresponding to $16 \, \mathrm{dB}$ of maximal squeezing.  The phase of the pump is chosen so that $\varphi=0$. The gray area indicates the band, where $\Gamma$ is real. For the spectrum of squeezing the first and the second approximations are rather far from the exact solution and only the third approximation is close to it. For the angle of squeezing the first approximation is significantly different from the exact solution, but already the second one is close to it.}
    \label{Fig::parameter_spectra_intermediate_gain}
	\end{figure}

In Fig.~\ref{Fig::parameter_spectra_intermediate_gain} we compare the frequency dependence of the spectrum of squeezing $s(\Omega)$ and the angle of squeezing $\psi_L(\Omega)$ for the exact solution, obtained in Sec.~\ref{Sec::Monochromatic_Pump_model}, and the three first orders of MA for the gain exponent $g=1.84$, corresponding to 16 dB of maximum squeezing obtained for perfect phase matching. Since frequency enters only via the phase-mismatch angle $\theta(\Omega)$, we use this angle as abscissa for the figures. We assume that the phase of the pump is chosen so that $\varphi=0$. We remind the reader that the angle in Fig.~\ref{Fig::parameter_spectra_intermediate_gain}b is the continuous version of the angle of squeezing, as discussed in Sec.~\ref{Sec::Monochromatic_Pump_model}.
		
Figure~\ref{Fig::parameter_spectra_intermediate_gain}a illustrates that the first-order Magnus approximation for the considered gain gives rather poor approximation for the exact solution. Moreover, the second-order approximation does not improve this difference, and only in the third-order the approximate solution approaches the exact one.

With the angle of squeezing, shown in Fig.~\ref{Fig::parameter_spectra_intermediate_gain}b, the situation is different: it is also rather far from the exact solution in the first approximation, but becomes much closer to the exact one already in the second-order approximation. Thus, for a monochromatic pump the even orders of the Magnus expansion mainly correct the angle of squeezing, while the odd orders mainly correct the degree of squeezing. We may conjecture that this behavior is applicable for higher orders as well and generally for the non-monochromatic pump.

It should be noted that a significant difference between the exact solution and the first-order Magnus approximation appears only for rather high values of the gain exponent $g$. For $g<1.15$, corresponding to squeezing below 10 dB, this difference is hardly visible. Thus, the first-order Magnus approximation can be effectively used in the regimes of the high-gain PDC where the maximal degree of squeezing is below a certain value. Above this limit the first-order approximation is not valid, and the higher orders of MA should be taken into account. We shall call a regime of PDC above this limit of squeezing {\it ultra-high-gain} PDC. The boundary for this regime depends on the acceptable error in the degree of squeezing. One possibility for giving such a definition is related to the distance between the first zeros of the degree of squeezing in the exact solution $r(\Omega)$ and its first-order approximation $r_1(\Omega)$, corresponding to the points $s(\Omega)=1$ in Fig.~\ref{Fig::parameter_spectra_intermediate_gain}. It follows from Eq.~\eqref{r} that the first zero of $r(\Omega)$ corresponds to the frequency where $B(\Omega)=0$. From Eq.~\eqref{AB} we find that this is the frequency where $\Gamma L=i\pi$ or $\theta=\sqrt{g^2+\pi^2}=\theta_0$. From Eq.~\eqref{r1} we obtain the first zero of $r_1(\Omega)$ as $\theta_1=\pi$. The relative distance can be defined as $d=(\theta_0-\theta_1)/\theta_1=\sqrt{(g/\pi)^2+1}-1$. For tolerable relative distance of 10\% we have $g\le1.44$, which corresponds to 12.5 dB of maximal squeezing. Thus, for PDC with monochromatic pump we can accept the value of 12.5 dB of maximal squeezing as the boundary between the high-gain and the ultra-high-gain regimes.

The numerical study of Ref.~\cite{ChristBMS2013} shows that for pulsed PDC this boundary is about 12 dB of squeezing, which is compatible with our analytical result.

\subsection{Convergence of the Magnus expansion}\label{Sec::Convergence}

In the previous subsection we have found the boundary value of the gain exponent $g$ above which the corrections from the higher-orders of the Magnus expansion are necessary. We can ask another question: what is the maximal value of $g$ for which the Magnus series converge and, therefore, make the Magnus expansion applicable? The question of convergence of the Magnus series has been studied in the literature, and it is generally known that the series converge if \cite{BlanesaCOR2009}
	\begin{align}
		\int\limits_{0}^{L}dz||\mathbf{F}(z)||_2<\pi
		\text{,}
		\label{Eq::Upper_Bound_Estimation}
	\end{align}
where $||M||_2$ stands for the so-called spectral norm of the matrix $M$ \cite{HornJohnson}. This norm can be calculated as the maximal singular value of $M$, or the square root of the maximal eigenvalue of the Hermitian matrix $M^\dagger M$. From Eq.~\eqref{Eq::Coupling_Matrix_F} we find that the maximal eigenvalue of $\mathbf{F}(z)^\dagger\mathbf{F}(z)$ is $|\sigma|^2$ and, therefore, $||\mathbf{F}(z)||_2=|\sigma|$. This value provides the upper bound of Eq.~\eqref{Eq::Upper_Bound_Estimation} as $g=\pi$, corresponding to squeezing of 27 dB. In practice, such degree of squeezing would require a very high coupling constant and would most probably invalidate the undepleted-pump approximation of our model. The record value of squeezing in the CW regime at present is 15 dB \cite{Vahlbruch2016}. Even if the limit of 27 dB for squeezing does not seem to be attainable experimentally in the near future, the theory allows us to use this value as the limit of convergence of the Magnus expansion.

\subsection{Homodyne detection of the down-converted light}	

Broadband squeezed light generated in the PDC is usually observed in the balanced homodyne detection scheme where a strong local oscillator field $\mathcal{E}_\mathrm{LO}^{(+)}(t,z)$ is mixed on a symmetric beam-splitter with the measured field $E^{(+)}(t,z)$ producing at the two outputs of the beam-splitter the fields
\begin{eqnarray}\label{E1}
E^{(+)}_1(t,L) = \frac1{\sqrt{2}}\left(E^{(+)}(t,L) + \mathcal{E}_\mathrm{LO}^{(+)}(t,L)\right),\\\label{E2}
E^{(+)}_2(t,L) = \frac1{\sqrt{2}}\left(E^{(+)}(t,L) -
\mathcal{E}_\mathrm{LO}^{(+)}(t,L)\right),
\end{eqnarray}
whose intensities are measured by two photodetectors with hight quantum efficiency. The observed quantity is the difference of photocurrents collected from two photodetectors, which we shall call simply photocurrent and denote $i(t)$. Its mean value is zero when the measured state is a squeezed vacuum, so we write the photocurrent fluctuation as $\delta i(t)=i(t)$. For detection of continuous-wave squeezing the local oscillator is chosen as a monochromatic wave $\mathcal{E}_\mathrm{LO}^{(+)}(t,z)=\mathcal{E}_0e^{i(k_0z-\omega_0z)}$, where $\mathcal{E}_0$ is a complex amplitude. The photon flux of the local oscillator is accepted to be much higher than that of the measured field, $|\mathcal{E}_0|^2\gg\langle E^{(-)}(t,L)E^{(+)}(t,L)\rangle$ and the quantum efficiency of both detectors is put to unity for simplicity. In this case the autocorrelation function of the photocurrent can be written as
\begin{eqnarray}\label{autocorr}
&&\langle \delta i(t)\delta i(t') \rangle = |\mathcal{E}_0|^2\delta(t-t')\\\nonumber
        &&+ \frac{|\mathcal{E}_0|^2}{(2\pi)^2} \int\langle : X(\Omega,L)X(\Omega',L):\rangle e^{-i\Omega t -i\Omega't'}d\Omega d\Omega',
\end{eqnarray}
where colons stand for normal ordering and $X(\Omega,L)$ is the measured quadrature, determined by the phase $\beta = \arg\left(\mathcal{E}_0\right)$ of the local oscillator:
\begin{eqnarray}\label{quadbeta}
     &&X(\Omega,L) = a(\Omega,L)e^{-i\beta} +a^{\dagger}(-\Omega,L)e^{i\beta} \\\nonumber
     && = X_1(\Omega,L)\cos\left[\psi_L(\Omega)-\beta\right] -X_2(\Omega,L)\sin\left[\psi_L(\Omega)-\beta\right].
\end{eqnarray}
The first term in the right hand side of Eq.~\eqref{autocorr} represents the shot noise, while the second one is proportional to the normally ordered autocorrelation function of the measured field quadrature. The shot noise level is determined by the mean sum of photocurrents of two detectors $\langle i_+\rangle = |\mathcal{E}_0|^2$.

\begin{figure}[ht]
		\includegraphics[width=8.6cm]{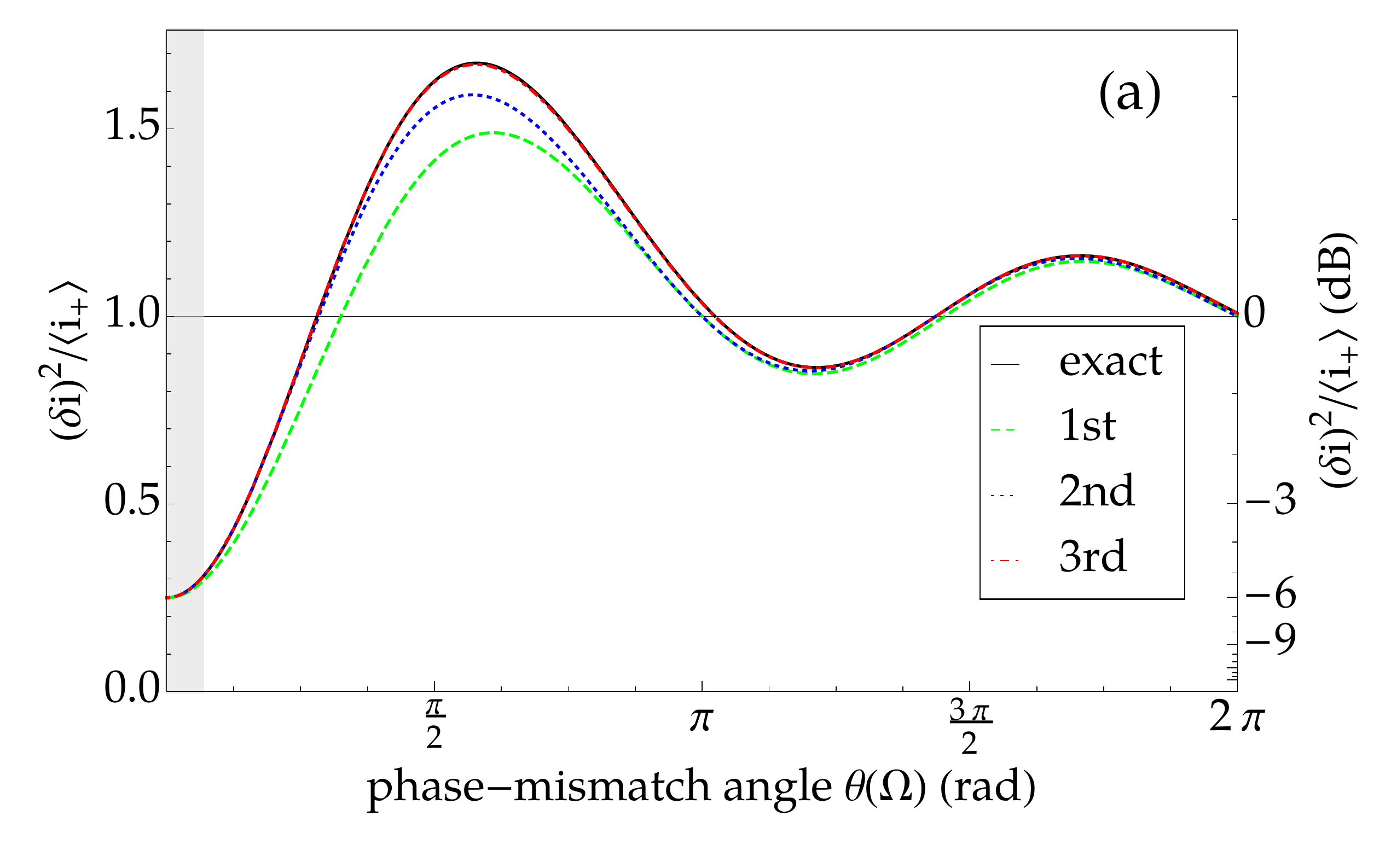}\\
		\includegraphics[width=8.6cm]{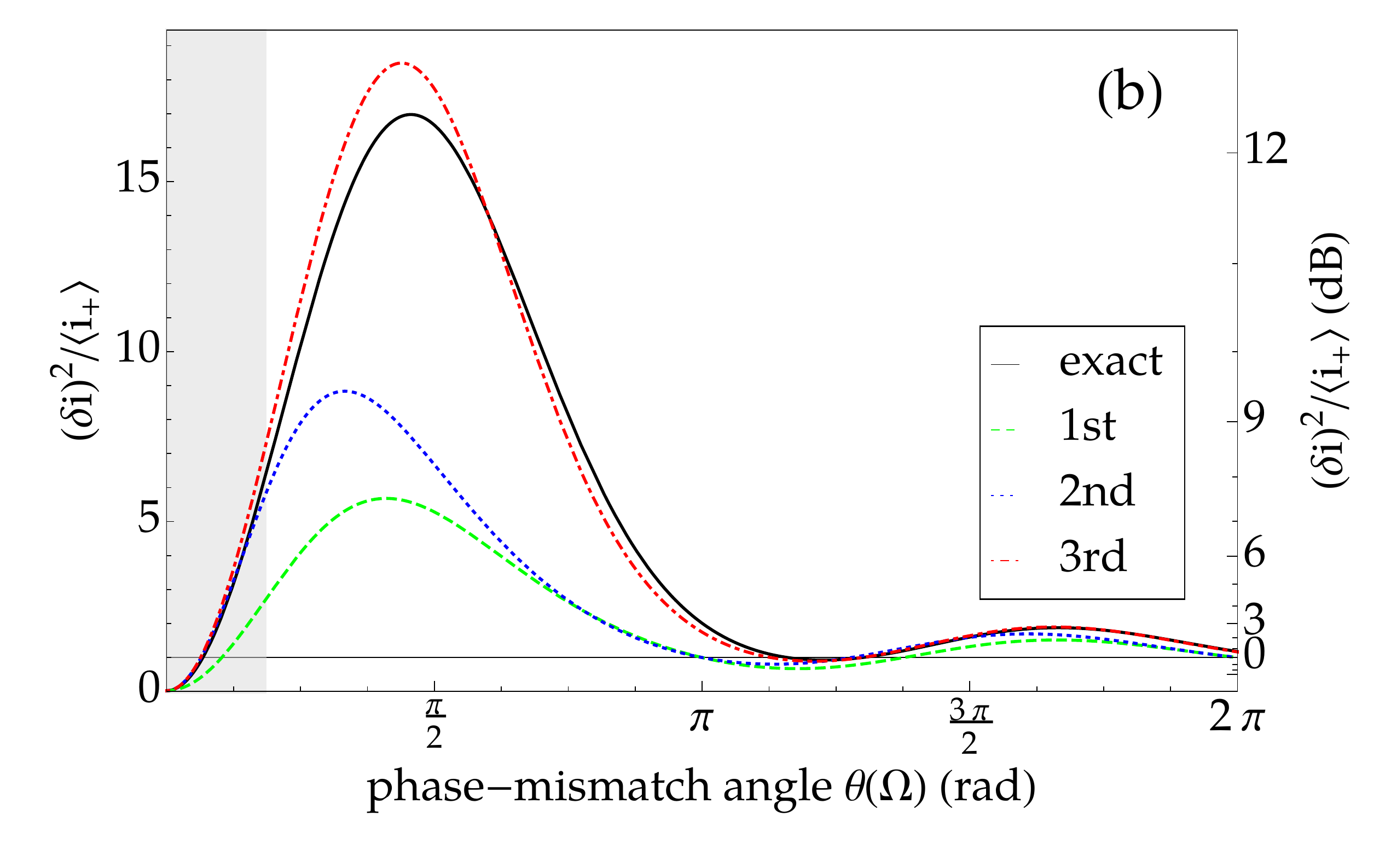}\\
		\caption{The normalized photocurrent noise spectrum for balanced homodyne detection of the down-converted light obtained from exact solution and three first orders of the MA. The gain exponents are chosen as (a) $g=0.7$, corresponding to $6 \, \mathrm{dB}$ of maximal squeezing  and (b) $g=1.84$, corresponding to $16 \, \mathrm{dB}$ of maximal squeezing. The gray area indicates the band, where $\Gamma$ is real.  At high squeezing only the third order approximation is satisfactory.}
    \label{Fig::Vacuum_fluctuations}
	\end{figure}

Defining the photocurrent spectral density $(\delta i)^2_\Omega$ as Fourier transform of the photocurrent autocorrelation function with respect to the time difference $t-t'$ and normalizing it to the shot noise level we obtain from Eqs.~\eqref{commutator}, \eqref{autocorr}, \eqref{quadbeta} and the requirement that for a quasi-stationary field \cite{MandelWolf} the correlator $\langle : X(\Omega,L)X(\Omega',L):\rangle$ is proportional to $\delta(\Omega+\Omega')$, the following expression:
	\begin{eqnarray}
		(\delta i)^2_\Omega/\langle i_+ \rangle &=& \frac1{2\pi}\int \langle X(\Omega,L)X(\Omega',L)\rangle d\Omega'
		\label{Eq::Homodyne}
		\text{,}
	\end{eqnarray}
which is valid for homodyne detection of any quasi-stationary field. For detection of squeezed vacuum, substituting Eq.~\eqref{corr} and \eqref{quadbeta} into Eq. \eqref{Eq::Homodyne}, we obtain
	\begin{eqnarray}
		(\delta i)^2_\Omega/\langle i_+ \rangle &=& \cos^2\left[\psi_L(\Omega)-\beta\right]e^{2r(\Omega)}\\\nonumber
        &+&\sin^2\left[\psi_L(\Omega)-\beta\right]e^{-2r(\Omega)}
		\label{Eq::Homo_Vac}
		\text{.}
	\end{eqnarray}
The effect of squeezing manifests itself as reduction of the fluctuations of photocurrent below the shot-noise level for particular choice of the phase of the local oscillator. It is, therefore, very interesting to see the effect of different orders of the MA on this experimentally observed quantity. We shall assume that $\beta$ can be chosen so that $\psi_L(\Omega)-\beta=\pi/2$ for the frequency $\Omega$ of perfect phase matching, where $\theta(\Omega)=0$, and squeezing is maximal. In Fig.~\ref{Fig::Vacuum_fluctuations} we present the normalized photocurrent noise spectrum for two different values of $g$ corresponding to moderate and high squeezing.

From Fig.~(\ref{Fig::Vacuum_fluctuations}) we observe that for moderate levels of squeezing shown in Fig.~(\ref{Fig::Vacuum_fluctuations})a the deviation of all three orders of MA from the exact solution remains tolerable, while for the high level of squeezing in Fig.~(\ref{Fig::Vacuum_fluctuations})b only the third-order MA give a tolerable approximation for the exact solution. The physical explanation of this effect is very simple: for high level of squeezing the photocurrent noise spectrum becomes much more sensitive to the errors in the squeezing angle in the corresponding order of the MA. This errors are responsible for the contribution into the photocurrent noise from the stretched component of the broadband squeezed state. Since this component is growing with the level of squeezing, the sensitivity to the errors increases accordingly.

\subsection{Dependence on the gain exponent $g$}\label{Sec::Nonlinear_Squeezing}

In the previous subsections we have considered the dependence of the degree of squeezing $r(\Omega)$, the angle of squeezing $\psi_L(\Omega)$, and the photocurrent noise spectrum on the frequency $\Omega$ for fixed gain exponent $g$. In this subsection we use a complementary approach and consider the dependence of the degree of squeezing against the gain exponent $g$, $r(g)$, for a fixed frequency $\Omega$. The gain exponent $g$ is proportional to the product of the pump amplitude, the crystal length and its nonlinear susceptibility. Thus, the analysis of this section can be understood as a study of influence of these three physical quantities on the validity of the various orders of the MA. The crystal length $L$ affects the eigenmode parameters also via the phase mismatch angle $\theta(\Omega)$. It means that the the dependence $r(g)$ shows the influence of the crystal thickness not at fixed frequency $\Omega$ but rather at fixed phase mismatch angle $\theta(\Omega)$.
	
The simplest case is one for the frequency of the perfect phase matching, $\Delta(\Omega)=0$, where we have for the exact solution and for all orders of MA, $r(g)=g$, i.~e.~a linear dependence on $g$ and, therefore, on the pump amplitude. It is remarkable, that this linearity is preserved in the first-order MA, as follows from from Eq.~\eqref{r1}.

\begin{figure}[ht]
		\includegraphics[width=8.6cm]{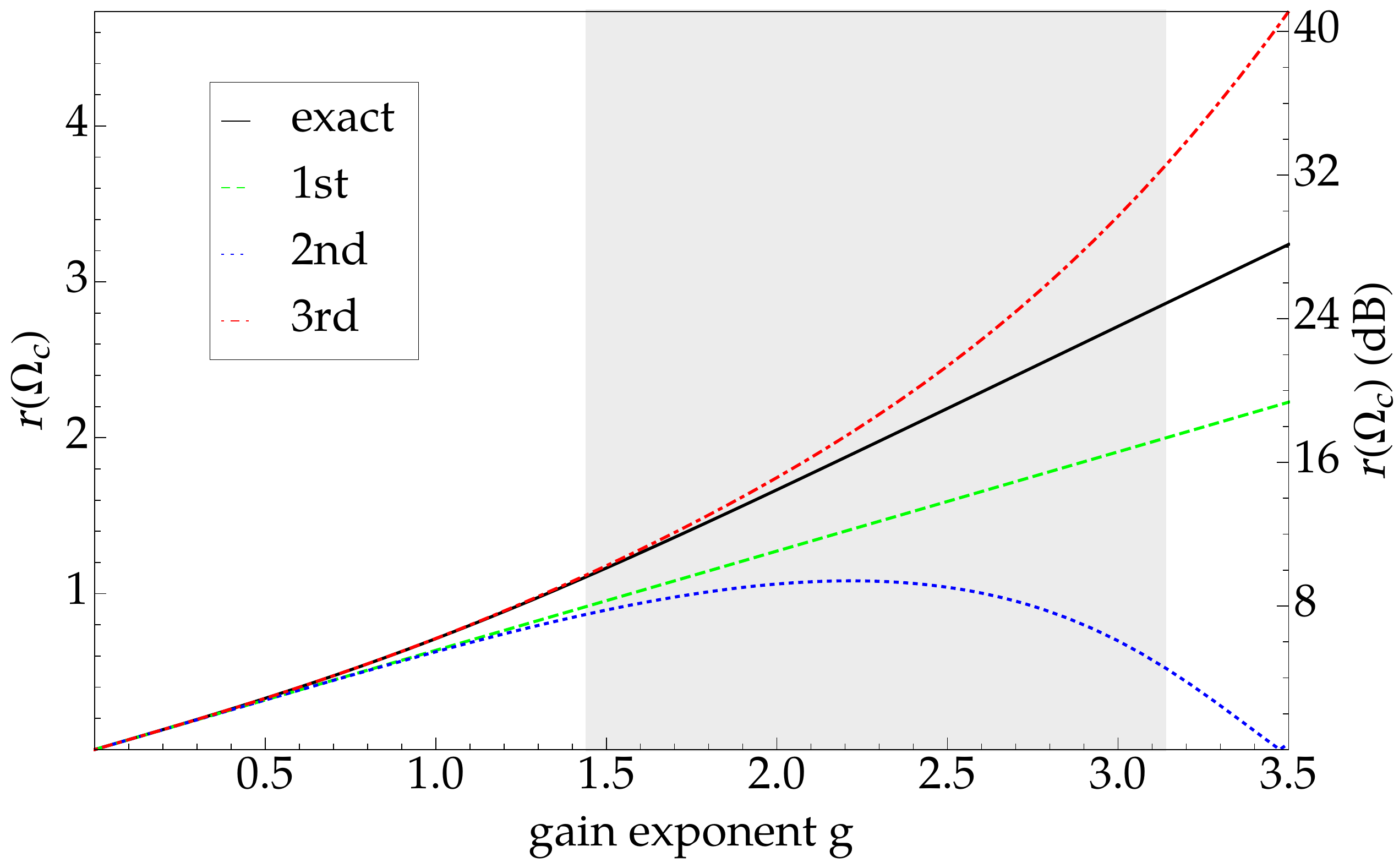}\\
		\caption{The degree of squeezing $r(\Omega_c)$ as function of gain exponent $g$ for $\Delta(\Omega_c)\neq 0$. The frequency $\Omega_c$ is chosen so that $\Delta(\Omega_c)L=\pi$. Four curves correspond to the exact solution $r$ (black, solid), the first-order MA $r_1$ (green, solid), the second-order MA $r_2$ (blue, dashed), and the third-order MA $r_3$ (red, dotted). The gray area indicates the region of the ultra-high gain, from $g=1.44$ to $g=\pi$ (12.5 to 27 dB of squeezing). Above $g=\pi$ the convergence of the Magnus expansion is not guaranteed.}\label{Fig::Nonlinear_squeezing}
	\end{figure}

In Fig.~\ref{Fig::Nonlinear_squeezing} we present the gain dependence of the degree of squeezing for non-zero phase mismatch, $\Delta(\Omega)\neq 0$. One can observe a nonlinear dependence of $r(g)$ against $g$ in the exact solution and a linear one in the first-order MA. Since the difference between the first-order MA and the exact solution is negligible for $g$ below the boundary of the ultra-high-gain, we conclude that deviations from linearity in the dependence of $r(g)$ can serve as a signature of the ultra-high-gain regime. One can also appreciate that the third-order MA improves the conversion towards the exact solution as compared with the the second-order MA in the gray area. Above the value of $g=\pi$ the convergence of the Magnus expansion is not guaranteed.
	
The dependence of the degree of squeezing on $g$ shown in Fig.~\ref{Fig::Nonlinear_squeezing} can be easily measured experimentally, since the gain exponent $g$ is proportional to the amplitude of the pump wave. A deviation from the linear dependence can be observed as difference of the parametric gain for non-zero phase mismatch from the behavior given by $\mu\sinh(\nu E_p)$, where $E_p$ is the pump amplitude, and $\mu$ and $\nu$ are some fitting parameters. Let us mention here that for aperiodically poled crystals this dependence is different even below the ultra-high-gain regime, and has been recently observed in the experiment~\cite{Chekhova2018}.

For better understanding the dependence of the degree of squeezing $r(\Omega)$ and its respective $\ell$-order Magnus approximations $r_\ell(\Omega)$ on $g$, we perform the Taylor expansions of $r_\ell(\Omega)$ in $g$:
	\begin{align}
		r_\ell(\Omega)&= \sum_{k=1}^\infty r_\ell^{[k]}(\Omega)\frac{g^k}{k!}
		\label{Eq::r_parameter_Taylor_expansion}
		\text{.}
	\end{align}

The analytical expressions for the Taylor coefficients in Eq.~\eqref{Eq::r_parameter_Taylor_expansion} up to the 4-th order are given in Tab.~\ref{Tab::Taylor_Coeff_for_r}.
	\begin{table}[ht]
\caption{The Taylor coefficients for the degree of squeezing  $r(\Omega)$ and its Magnus approximations up to $4$th order in the gain exponent $g$. We remind that $\theta(\Omega)=\Delta(\Omega)L/2$.
		}\label{Tab::Taylor_Coeff_for_r}
		\begin{tabular}{ccccc}\toprule
		$k$ & $r^{[k]}(\Omega)$ & $r^{[k]}_1(\Omega)$ & $r^{[k]}_2(\Omega)$	& $r^{[k]}_3(\Omega)$ \\\colrule
		$1$ & $j_0(\theta)$		& $j_0(\theta)$	      &	$j_0(\theta)$	    &	$j_0(\theta)$	\\
		$2$ &  0			    &	0		          &	0		            &	0				\\
		$3$ & $j_0(\theta)-j_0^3(\theta)+j_2(\theta)$ & 0 &	0  & $j_0(\theta)-j_0^3(\theta)+j_2(\theta)$ \\
        $4$ &  0           	    & 0                   &	0		            &	0                  \\\botrule
		\end{tabular}
		
	\end{table}

As follows from this Table, the correct value of the first-order Taylor coefficient $r^{[1]}_1(\Omega)=r^{[1]}(\Omega)$ appears in the first-order MA. The second-order Taylor coefficient for $r(\Omega)$ vanishes, since the latter is an odd function of $g$. As a result, the second-order MA makes no correction to the degree of squeezing in the second order of $g$. This observation corroborates the result of Ref.~\cite{QuesadaS2014} where the authors have predicted that for PDC with vacuum input the second order MA provides no correction in the second order in $g$. The third-order MA gives the correct value of the third-order Taylor coefficient $r^{[3]}_3(\Omega)=r^{[3]}(\Omega)$. We can conjecture that the correct value for the $k$th Taylor coefficient appears in the $k$th order of the Magnus expansion.

A similar decomposition can be written for the angle of squeezing $\psi_L(\Omega)$:
    \begin{align}
		\psi_{L,\ell}(\Omega)&= \sum_{k=0}^\infty \psi_{L,\ell}^{[k]}(\Omega)\frac{g^k}{k!}
		\label{Eq::psi_parameter_Taylor_expansion}
		\text{,}
	\end{align}
with the corresponding coefficients shown in Tab.~\ref{Tab::Taylor_Coeff_for_psiL}.

\begin{table}[h!]
\caption{The Taylor coefficients for the angle of squeezing  $\psi_L(\Omega)$ and its approximations up to $3$-rd order in the gain exponent $g$. We have introduced a shortcut $\zeta(\theta)=\frac12\left(\sin (\theta ) j_0(\theta )-\cos (\theta ) j_1(\theta )\right)$.
		}\label{Tab::Taylor_Coeff_for_psiL}
		\begin{tabular*}{\columnwidth}{ccccc}\toprule
		\hspace{0.3cm} $k$ \hspace{0.3cm} & \hspace{0.3cm}$\psi^{[k]}_{L}(\Omega)$\hspace{0.3cm} & \hspace{0.3cm}$\psi_{L,1}^{[k]}(\Omega)$\hspace{0.3cm} & \hspace{0.3cm}$\psi_{L,2}^{[k]}(\Omega)$\hspace{0.3cm}	& \hspace{0.3cm}$\psi_{L,3}^{[k]}(\Omega)$\hspace{0.3cm} \\\colrule
		$0$ & $\frac12\left(\varphi-\theta\right)$		&  $\frac12\left(\varphi-\theta\right)$	      &	 $\frac12\left(\varphi-\theta\right)$	    &	 $\frac12\left(\varphi-\theta\right)$	\\
		$1$ &  0			    &	0		          &	0		            &	0				\\
		$2$ & $\zeta(\theta)$  & $0$ & $\zeta(\theta)$ & $\zeta(\theta)$\\
        $3$ &  0           	    & 0                   &	0		            &	0                  \\
        \botrule
		\end{tabular*}
		
	\end{table}
From Table II we conclude that for the angle of squeezing the correct value of the $k$th Taylor coefficient is given by the $k$th and above orders of the MA, at least for the first 4 orders. We can conjecture that this dependence holds as well for the higher orders of the Magnus expansion.

\section{Conclusion}\label{Sec::Summary}

	We have applied the Bloch-Messiah decomposition to the process of the type-I parametric down-conversion in a second-order nonlinear crystal with monochromatic pump. Using an exact solution known for this process, we have evaluated the four real parameter characterizing the Bloch-Messiah decomposition and have introduced the squeezing eigenmodes which are in a single-mode squeezed state and, therefore, are statistically independent. We have shown that for the monochromatic pump the eigenmodes are bichromatic and are parameterized by two angles. Next, we have applied the Magnus expansion to the quantum-mechanical evolution operator of this system and obtained analytic expressions for the first three orders of the Magnus approximation. We have shown that above certain degree of squeezing corrections to the first-order MA are necessary, and have introduced a boundary value of the parametric gain exponent $g=1.44$, corresponding to 12.5 dB of squeezing, as a boundary for the ultra-high-gain regime of PDC. We have demonstrated that for squeezing as high as 16 dB the third-order MA provides a very good approximation of the broadband squeezed squeezed light generated in this process.

	We have shown that a nonlinear dependence of the degree of squeezing $r(g)$ for non-zero phase mismatch can serve as a signature of the ultra-high-gain regime of PDC, a result which can be verified experimentally. We have also demonstrated that the photocurrent noise spectrum in the balanced homodyne detection of broadband squeezed light is very sensitive to the errors in the angle of squeezing in the respective Magnus approximations for ultra-high-gain regime. Our results confirm that the first-order MA, used in several previous publications, can be trusted for moderate squeezing, and provide the level of squeezing for which the higher-orders corrections are necessary.
	
	We expect that many of our results, obtained for the monochromatic pump, will remain valid for narrow-band non-monochromatic pump. This case will be studied in the subsequent publication.

\section*{Acknowledgments}

This work was supported by the European Union's Horizon 2020 research and innovation programme under grant agreement No 665148 (QCUMbER).

\end{document}